\documentclass[12pt,preprint]{aastex}

\shorttitle{Multi-band Variability Analysis of \object{Mrk 421}}
\shortauthors{H. Z. Li et al.}

\begin{document}


\title{Multi-band Variability Analysis of \object{Mrk 421}}

\author{H. Z. Li\altaffilmark{1}, Y. G. Jiang\altaffilmark{2,\dag}, D. F. Guo\altaffilmark{2}, X. Chen\altaffilmark{2}, T. F. Yi\altaffilmark{3}}
\altaffiltext{1}{Physics Department, Yuxi Normal University, Yuxi, Yunnan,
653100, China}
\altaffiltext{2}{Shandong Provincial Key Laboratory of Optical Astronomy and Solar-Terrestrial Environment, Institute of Space Sciences, Shandong University,Weihai, 264209, China}
\altaffiltext{3}{Physics Department, Yunan Normal University, Kunming, Yunnan, 650092, China, yitingfeng98@163.com}
\altaffiltext{\dag}{Corresponding author: jiangyg@sdu.edu.cn}





%


\begin{abstract}
We have assembled the historical variability
data of \object{Mrk 421} at radio 15 GHz, X-ray and $\gamma$-ray bands, spanning about 6.3, 10.3 and 7.5 yr, respectively. We analyzed the variability by using three methods. The results indicated that there is a period of $287.6\pm4.4$ days for 15 GHz, $309.5\pm5.8$ days for X-ray and $283.4\pm4.7$ days for $\gamma$-ray, respectively. This period can be reasonably explained by the nonballistic helical motion of the emitting material. The correlation analysis suggested that the \textbf{variabilities} of radio 15 GHz, X-ray and $\gamma$-ray
are remarkable correlated, and the emission of radio 15 GHz lags
behind that of X-ray, and the X-ray flux lags
behind the $\gamma$-ray. This suggests that the
$\gamma$-ray derives from inverse Compton (IC) scattering of the synchrotron photons, supporting the synchrotron self-Compton (SSC) model. Moreover, the time delay between different wavebands could be explained by the shock-in-jet models, in which a moving emission region produces the radio to $\gamma$-ray\textbf{ activity}, 
implying that the emission region of $\gamma$-ray is closer to the center than ones of X-ray and radio emission.

\end{abstract}





\section{Introduction}
Blazars are a subclass of active galactic nuclei
(AGNs) with strong \textbf{variabilities}. 
The variability
analysis is the most powerful tool to probe the emission mechanism and processes \citep{Chan14,Sill88,Lain99}.
Based on the variability timescales, the \textbf{variabilities} can be roughly divided into three classes: the intra-day variability (IDV) with the
timescale from several minutes to a day, \textbf{the} short term variability (STV) with the timescale from several days to several months, and \textbf{the} long term variability (LTV) on the
timescale of years \citep{Gupt08,Gaur12}.
\textbf{Blazar} variabilities have been well studied by many authors \citep[e.g.][]{Chen14,Li06,Li09,Li10,Li15,Poon09,Qian07,Rait03,Sill88,Xie08,Zhan08}.

\object{Mrk 421} (B2 1101+384) is the first extragalactic source detected at TeV \textbf{energy}, and is one of the nearest and brightest BL Lac \textbf{object} with a redshift $z=0.031$ \citep{Lico12,Punc92}. It is classified as \textbf{the} high energy peaked BL Lac \textbf{object} (HBL), because the synchrotron hump peaks at soft X-rays. In the optical band, strong variability was observed for this object \citep{Gaur12,Liu97,Mill75}. \citet{Xie88} discovered that the optical brightness changed 1.4 $mag$ in $2.5$ hr. \citet{Stei76} found a LTV with 4.6 $mag$.  In X-ray and $\gamma$-ray bands, the source is almost always active with \textbf{a} big eruption on a timescale about two years \citep{Bart11,Cui04,Gaur12,Tluc10}. Moreover, a major outburst always accompanies with many quick flares \citep{Bart11}.
In the spring-summer 2006, \citet{Tram09} found that the X-ray flux of \object{Mrk 421} reached its highest \textbf{record}. \textbf{In 2006 April, \object{Mrk 421} underwent a large flare and the X-ray flux was variable \citep{Ushi09}.} In 2010 January and February, big X-ray outbursts were observed from the object \citep{Isob10}. X-ray and TeV outbursts were also reported for the source by \citet{Kerr95} and \citet{Taka94,Taka95} in 1994 and 1995, respectively.

Different \textbf{timescale variabilities} of \object{Mrk 421} have been intensively investigated \citep{Blas13}. A radio flare at 43 GHz during 2011 on a timescale about $\triangle t\sim150$ days was reported \textbf{in the VLBI core} by \citet{Blas13}.
\citet{Race13} studied the multiwavelength data of \object{Mrk 421}, and the results suggest that the radio emission is consistent with $\gamma$-ray on timescales from 1 day to 6 months. For infrared emission, IDV and STV were found by \citet{Gupt04} for the source. Variability on timescales from 32 to 256 days in X-ray and TeV bands was discovered by \citet{Smit08}. By using the structure function, \citet{Hora09} found that the variability timescales are $\triangle t\sim5$  and $8$ days for the Proportional Counter Array (PCA) data, $\triangle t\sim20$  and $25$ days for the All Sky Monitor (ASM) data, $\triangle t\sim55$  and $71$ days for the Burst Alert Telescope (BAT) data, and $\triangle t\sim40$  and $60$ days for the optical data. \citet{Bart11} reported that \object{Mrk 421} would undergo a major outburst about once every two years in X-ray and $\gamma$-ray bands, and the outburst usually lasts several months. By using the Jurkevich method, \citet{Liu97} found that there are two possible
periods with timescales of $23.1\pm1.1$ and $15.3\pm0.7$ years in the B band of \object{Mrk 421}. \citet{Chen14} investigated the B and V bands historical variability data, and a possible
period of about 1.36 years was found.
These previous results suggest that various quasi-periodic timescales exist in the variability of \object{Mrk 421} at multiple wavebands.

The \textbf{correlations} among different wavelengths of \object{Mrk 421} have been intensively studied. The correlation analysis of X-ray and $\gamma$-ray variability suggests that there is a significant correlation between them \citep{Balo13,Acci14,Acci11,Abdo14,Foss08,Hora09,Smit08,Ushi10,Bart11,Blaz05}.
 \citet{Gaur12} investigated the correlation between the optical and X-ray bands, and found that the optical variability lags behind the X-ray variability for $9.5\pm2$ days.
\citet{Max14} investigated
the relationship between the radio and $\gamma$-ray emission of \object{Mrk 421}, and found that the $\gamma$-ray variability is well correlated with radio variability lagging for $40\pm9$ days. \textbf{\citet{Rich13} found that the 15 GHz and LAT $\gamma$-ray light curves are well correlated around the 2012 flares, with 15 GHz lagging 40 days behind.} Moreover, \citet{Lico14} suggested that the radio and $\gamma$-ray light curves of \object{Mrk 421} are correlated with zero delay.

Obviously, \object{Mrk 421} is \textbf{a} very active and highly
variable source with variable timescales. In this work, we extracted the radio 15 GHz, X-ray and $\gamma$-ray
light curves from the published database. Both the variability and the correlations among different wavelengths are analyzed in detail.
This paper is arranged as follows: the data assembling is described in Sect. 2; the periodic analysis methods and the results are given in Sect. 3; the correlation analysis is given in Sect. 4;
in the last section, we give discussion and conclusions.

\section{Observation Data and Variability Analysis of the Light Curves}
We presented the variability data of \object{Mrk 421} at radio 15 GHz, X-ray (0.3-10 keV) and
$\gamma$-ray (0.1-300 GeV). The 15 GHz data of \object{Mrk 421} were taken from the 40 m Telescope at the Owens Valley Radio Observatory (OVRO)\footnote[1]{http://www.astro.caltech.edu/ovroblazars/data/data.php?page=data\_return\& source=J1104+3812}. The objects measured by OVRO at 15 GHz are same as ones observed by Fermi $\gamma$-ray, which can promote the understanding of the emission mechanisms of blazars. The light curve of 15 GHz spans about 6.3 yr from January 7th, 2008 to May 14th, 2014.

The data of X-ray (0.3-10 keV) were obtained from the database of the X-ray
Telescope (XRT) in Swift, and were processed using the web-based tool\footnote[2]{http://www.swift.ac.uk/user\_objects/} \citep{Evan09}. Using this software, the light curves of \object{Mrk 421} were obtained using the observation bin and 3-$\sigma$ upper limits. The overall light curves contain all Swift-XRT observations in Photon Counting (PC) and Windowed Timing (WT) modes \citep{Burr05,Evan09} from February 28th, 2005 to June 24th, 2015.

The $\gamma$-ray data ($0.1-300$ GeV) are measured by Fermi LAT (The Large Area Telescope on the Fermi Gamma Ray Space Telescope spacecraft), and the weekly variability data
are taken from the Fermi light
curves\footnote[3]{http://fermi.gsfc.nasa.gov/ssc/data/access/lat/msl\_lc/}.
The $\gamma$-ray data used in this paper were observed during \textbf{August} 2th, 2008 to August 18th 2015.

The light curves are displayed in Figure 1. Figure 1 shows that \object{Mrk 421} is a highly active source, with a variability index $V_{15 GHz}=0.50$, $V_{X}=0.98$ and $V_{\gamma}=0.87$. The variability
index is defined by the formula \citep{Fan02,Li09}
\begin{equation}
 V=\frac{F_{max}-F_{min}}{F_{max}+F_{min}},
\label{eq:Lebseque V}
\end{equation}
where $F_{max}$ and $F_{min}$ is the maximal and minimal flux, respectively.


\section{Analysis Method and Periodicity Results}

\subsection{The results of the Lomb-Scargle periodogram}

The Lomb-Scargle periodogram \citep{Lomb76,Scar82} is a traditional method, and can be applied to analyze the periodicity in the nonequispaced variability data. 
The periodogram is a function of circular frequency $\omega$, and is defined by the formula

\begin{equation}
 P_{X}(\omega)\equiv \frac{1}{2}\{\frac{[\Sigma _{i}X(t_{i})cos\omega(t_{i}-\tau)]^{2}}{\Sigma_{i}cos^{2}\omega (t_{i}-\tau)}+\frac{[\Sigma _{i}X(t_{i})sin\omega(t_{i}-\tau)]^{2}}{\Sigma_{i}sin^{2}\omega
 (t_{i}-\tau)}\},
\label{eq:LebsequeLS1}
\end{equation}
where $X(t_{i})$ ($i=0, 1...,N_{0}$) is a \textbf{time} series. $\tau$ is calculated by the equation
\begin{equation}
 \tau=\frac{1}{2\omega} tan^{-1}[\frac{\Sigma _{i}sin2\omega t_{i}}{\Sigma _{i}cos2\omega
 t_{i}}],
\label{eq:LebsequeLS2}
\end{equation}
where $\omega=2\pi\nu$. Thus, the periodogram is a function of
frequency $\nu$.
For a true signal $X(t_{i})$, the power in $P_{X}(\omega)$ would present a peak, or the power of a purely noise signal would be a
exponential distribution. For a power level $z$, the False Alarm Probability (FAP) is calculated by \citep{Scar82,Pres94}
\begin{equation}
 p(>z)\approx N\cdot exp(-z),
\label{eq:LebsequeIp}
\end{equation}
where $N$ is the number
of data point \citep{Pres94}.
%
%

The result of the Lomb-Scargle periodogram is shown in
Figure 2. The (a), (b) and (c) panel of Figure 2 give the results of 15 GHz, X-ray and $\gamma$-ray, respectively. The (a) panel shows that possible periods are 181.9, 206.1, 237.8 and 289.9 days in the light curves of 15 GHz. The FAP of the results is computed using the equation 4, and the red line in the (a) panel of Figure 2 \textbf{denotes} the FAP levels of 0.01. The (a) panel suggests that the FAP levels of the four possible timescales are smaller than 0.01 (FAP$<0.01$), which means that the confidence levels of all the timescale are larger than $99\%$.
The (b) and (c) panel of Figure 2 implies that \textbf{the} possible \textbf{periods are} 311.7 and 281.1 days in the X-ray and $\gamma$-ray light curves, respectively. The (b) and (c) panel also displays that the FAP levels of all timescales are smaller than 0.01.
The results of the X-ray and $\gamma$-ray \textbf{are} consistent with the period of 289.9 days in the 15 GHz light curve.

In order to test the reliability and the significance level of the possible detected peak signals in the power spectra, we also estimated the red noise significance level using the REDFIT38 software\footnote[4]{http://www.geo.uni-bremen.de/geomod/staff/mschulz/\#software2} \citep{Schu02}, which fits a first-order autoregressive process to the time series to estimate the red-noise spectrum. The top, middle and bottom panel of Figure 3 give 15 GHz, X-ray and $\gamma$-ray results calculated by REDFIT38 software. Figure 3 shows that the variability timescales $f=0.003443$ ($P=290.4$ days) with significance level $p>80\%$ for 15 GHz,  $f=0.003318$ ($P=301.4$ days) with significance level $p>99\%$ for X-ray, and  $f=0.003489$ ($P=286.6$ days) with significance level $p>99\%$ for $\gamma$-ray.

\subsection{The Jurkevich method and results}
Based on the standard deviation, the Jurkevich method \citep{Jurk71} is a very useful technology to seek the variability signal from the nonequispaced variability data. This method \textbf{tests} a range of possible \textbf{signals}. According to a possible signal, the data are folded and split into $m$ terms.
%
For the $lth$ term of a data series ${x_{i}}$, the parameters $\overline{x}_{l}=\frac{1}{m_{l}}\sum_{i=1}^{m_{l}} x_{i}$ and
$V^{2}_{l}=\sum_{i=1}^{m_{l}} x^{2}_{i}-m_{l}\overline{x}^{2}_{l}$ are calculated. Then, the sum of the squared
deviations of $m$ terms $V^{ 2}_{ m}=\sum_{l=1}^{m} V^{2}_{l}$ is also calculated. If the possible signal tested is a genuine one, the value of $V^{ 2}_{ m}$ would attain its minimum value \citep{Jurk71}. 

The (a), (b) and (c) panel of Figure 4 give the results of 15 GHz, X-ray and $\gamma$-ray light curves of Figure 1 from the Jurkevich
method, respectively. The (a) panel of Figure 4 shows that there is \textbf{a} clear least value of $V_{m}^{2}$ plot at the timescale of
$289\pm11.5$ days, indicating a possible period in the trial.
The red line in the (a) panel is the FAP level of 0.01, which gives a quantitative criterion for the detection of a minimum \citep{Fan10,Horn86}. The FAP was computed by a simple Monte Carlo method with N=10000. The (a) panel suggests that the FAP level of the possible timescales of $289\pm11.5$ days is smaller than 0.01 (FAP$<0.01$).

The (b) panel of
Figure 4 indicates that the timescales are $315\pm5$ and
$630\pm51.5$ days for X-ray light curves. Considering that 630 days is two times of 315 days, we infer that $315\pm5$ days is the possible period of X-ray light curve for \object{Mrk 421}, which is in agreement with the results of 15 GHz. From the (c) panel of Figure 4, period of 278, 571 and 833 days are confirmed. Since the timescale of 278 days is about one half and third of the timescales of 571 and 833 days, respectively, we think that 278 days is the possible period of $\gamma$-ray variability for \object{Mrk 421}. The (b) and (c) panel also show that the FAP levels of the results of X-ray and $\gamma$-ray are smaller than 0.01. Figure 4 implies that the variability timescales derived by the Jurkevich method using 15 GHz, X-ray and $\gamma$-ray variability data of \object{Mrk 421} are consistent.

\subsection{The discrete correlation function (DCF) method and results}

The discrete correlation function (DCF) \citep{Edel88} is a powerful technique to search for the correlation and time delay between two times series.
For a single times series, this method can be applied to seek the variability timescales in the time series. If there is a true signal on timescale of $T_{pe}$ in the variability time series, the DCF would attain its
maximum value at time delay $\tau=0$ and $\tau=T_{pe}$.
Namely, it will show a peak at the lags $\tau=0$ and $\tau=T_{pe}$.

The values of the DCF of each pair of data
$(a_{i},b_{j})$ can be calculated as follows:
At first, we compute the set of unbinned discrete
correlations (UDCF) of all pairs of data $(a_{i},b_{j})$ by the formula,
\begin{equation}
 UDCF_{ij}=\frac{(a_{i}-\bar{a})(b_{j}-\bar{b})}{\sigma_{a}\sigma_{b}},
\label{eq:LebsequeDCFI}
\end{equation}
where $\bar{a}$ and $\bar{b}$ are the average of the data sets,
$\sigma_{a}$ and $\sigma_{b}$ are the corresponding standard
deviations. Each pair of data can be related to the pairwise
lag $\triangle t_{ij}=t_{j}-t_{i}$.
Then, the $DCF(\tau)$ can be calculated by the formula,

\begin{equation}
 DCF({\tau})=\frac{1}{M}\Sigma UDCF_{ij}(\tau),
\label{eq:LebsequeDCFII}
\end{equation}
where $M$ is the number of pairs for which $\tau-\triangle\tau /2
\leq \triangle t_{ij}<\tau + \triangle \tau /2$, and $\tau$ is the
time lag. The standard error for each time lag $\tau$ is
\begin{equation}
 \sigma(\tau)=\frac{1}{M-1}\{\Sigma[UDCF_{ij}-DCF(\tau)]^{2}\}^{1/2}.
\label{eq:LebsequeDCFIII}
\end{equation}

The analytical results of DCF are displayed in Figure 5. Figure 5 suggests that the periods $T_{pe}$ derived by the DCF method are 186 and 281 days for 15 GHz light curves; 67, 202, and 310 days for the X-ray light curves; 86 and 288 days for the $\gamma$-ray light curves.

\subsection{Periodicity results}
Figure 2-5 and Table 1 give the results of the light curve of \object{Mrk 421} at
three different bands with three kinds of period analysis techniques. The foregoing analyses and Table 1 show that a certain result, 281.0-290.4 days for 15GHz variability dataset, 301.4-315.0 days for X-ray variability dataset, 278-288 days for $\gamma$-ray variability dataset were confirmed by three kinds of period analysis methods. This also shows that the timescales $278-315$ days obtained from three different wavelengths is consistent. From table 1, one can also find the other timescales, such as 181.9, 206.1 237.8, 186 days for 15 GHz, 67, 202 days for X-ray, and 86 days for $\gamma$-ray, only appear in the result of a single method. Therefore, these results should be excluded, and more observation are needed to confirm them.
Moreover, the Figure 1 shows that the light curves of \object{Mrk 421} are dominated by the big 2012 flare, which is most likely not a periodic event. In order to study the extent of the flare influence on the variability timescales, we also subtract the 2012 flare from the data. \textbf{The excluded period, a shaded area overlaid on Figure 1, is from MJD 56178 to 56230 for 15 GHz, and from MJD 56109 to 56188 for $\gamma$-ray band. Moreover, the X-ray data of the 2012 flare is lost (see Figure 1, a gap at the same time interval). Then, we analyze the new light curves removing the 2012 flare data using the Lomb-Scargle periodogram and Jurkevich methods, and Figure 6 gives the results.} From Figure 6, one can find that the timescales is 279-288 days with FAP level smaller than 0.01, which is consistent with the results obtained from the data containing the 2012 flare data. Therefore, the impact of the \textbf{episodic} flares on the variability timescale is very small.
Thus, the foregoing analyses and Table 1 implies that there is a true period on timescale of $287.6\pm4.4$ days (the average of 289.9, 290.4, 289, and 281 in the results of 15 GHz) in 15 GHz light curves, $309.5\pm5.8$ days (the average of 311.7, 301.4, 315, and 310 in the results of X-ray) in X-ray light curves, $283.4\pm4.7$ days (the average of 281.1, 286.6, 278, and 288 in the results of $\gamma$-ray) in $\gamma$-ray light curves. The variability periodicity of three different wavelengths is \textbf{broadly} consistent with each other.

\section{Correlation Analysis}

The correlations in the variability can improve our understanding of the emission mechanisms of blazars.
The DCF method can be used to calculate the correlation and time lag between
different light curves.
Thus, we will be looking for the correlations between the variability of 15 GHz, X-ray and $\gamma$-ray using the DCF method.
The Figure 7 gives the correlation result calculated by the DCF method. From  Figure 7, one can find that there is a
significant peak at the timescale of $\tau \simeq 52$ days in the X-ray /
15 GHz panel; $\tau \simeq 55$ days in the $\gamma$-ray /
15 GHz panel, and $\tau \simeq 2$ days in the $\gamma$-ray /
X-ray panel. This suggests that the emission of radio 15 GHz lags behind that of X-ray, and the X-ray variability lags the $\gamma$-ray variability.
To assess the significance of the correlation, we followed the approach presented by \citet{Chat08}. Firstly, we simulated 10000 light
curves using the algorithm recommended by \citet{Timm95} with a power law slope of 1.5 \citep{Lico14}. The mean and standard deviation of the simulated
curves is same with ones of the observed light curves.  Secondly, \textbf{based on} the original light curve, we resampled the simulated
curves. Thirdly, we analyzed the correlation of the simulated light curves, and identified the peak in each of the correlations of the simulated
curves. Finally, we compared the peak values obtained from the simulated curves with the ones computed from the the observed light curves. The results shows the significance levels of the correlation between the X-ray and 15 GHz, and $\gamma$-ray and 15 GHz, and $\gamma$-ray and X-ray are 85.82\%, 100\%, and 97.26\%, respectively.
This implies that the emissions at 15 GHz, X-ray and
$\gamma$-ray band are \textbf{well} correlated.
This agrees with the previous results obtained by many authors \citep[e.g.][]{Abdo14,Balo13,Bart11,Blaz05,Foss08,Gaur12,Lico14,Rich13,Smit08,Ushi10}, who found that there are clear \textbf{correlations} among different energy regimes, and the lower frequencies variations lag behind the higher frequencies variability.

\section{Discussion and Conclusions}
We presented the long term historical light curves of radio 15 GHz, X-ray and $\gamma$-ray
bands of \object{Mrk 421}, and analyzed the
period of the light curves in the three
different wavebands using various methods.
Possible period on timescales of $287.6\pm4.4$ days in 15 GHz light curves, $309.5\pm5.8$ days in X-ray light curves, $283.4\pm4.7$ days in $\gamma$-ray light curves were confirmed by three different methods. The period of $283.4-309.5$ days in 15 GHz, X-ray and $\gamma$-ray bands light curves are similar with the variability timescale of $256$ days reported by \citet{Smit08}.
The periodic variability may be associated
with some physical
time scales, e.g. the lasting time of big outbursts or a
total time intervals of many rapid flares close in time \citep{Kart07}.
Nevertheless, the light curves of 15 GHz, X-ray and $\gamma$-ray presented in Figure 1 \textbf{show} that there are no any obvious activities lasting for about 300
days. 
This shows that the variability on timescales of  $283.4-309.5$ days got in this paper
may be related to a periodic nature of the intrinsic variability.

For blazars, the emission is usually dominated by non-thermal emission from the relativistic jet, and the emission variability is caused by the
disturbance propagating down the jet.
The cyclic variation may be usually ascribed to the seasonally changing beaming effects caused by the periodic change of the visual angle, providing a
geometrical explanation for the blazar emission variability \citep{Brit09}. The visual angle changing over time suggests that the emitting material
moves along a curved track.
Many observational phenomena of \textbf{blazars} have been explained by geometrical origin \citep[e.g.][]{Osto04,Rait01,Brit09, Poon09,Li06,Li09,Li10,Li15,
Fuhr06,
Rait11}.
The changing viewing angle can be explained in the term of the bending jet, wiggling jet and the helical magnetic fields in jet \citep{Li15}. The supermassive binary black hole (SMBBH) model can offer \textbf{a} sufficient interpretation for the bending jet and wiggling jet \citep{Li15}.
Now, it is \textbf{expected} that SMBBH really exists in the center of
blazar \citep[e.g.][]{Bege80,Liu14}. The SMBBH model has been successfully applied to explain a number of observational phenomena
of blazar
\citep[e.g.][]{Chen14,Brit12,Li15,Valt11,Sill88,Tsal11}.
We assume that SMBBH exists in the center of \object{Mrk 421} and a helical magnetic field exists in the jet. \citet{Chen14} have interpreted the multiwavelength behaviour of \object{Mrk 421} using the SMBBH model.
Therefore, the
emitting material would move along a asymmetrical curved track generated by the jet with a precession and helical
magnetic field. This \textbf{leads} to a nonballistic helical motion for the
emitting material \citep{Brit10,Li15}. The periodicity
variability in the light curves of \object{Mrk 421} on the timescale  $283.4-309.5$ days can be
reasonably interpreted by the nonballistic helical motion of the
emitting material.

A multiwavelength cross-correlation analysis shows that there is a \textbf{well} correlation between the light curves of
the radio 15 GHz, X-ray and $\gamma$-ray waveband, which
is in agreement with the conclusions obtained in literatures \citep[e.g.][]{Gaur12,Balo13,Abdo14,Foss08,Smit08,Ushi10,Bart11,Blaz05, Rich13}.
The correlation between the radio, X-ray and
$\gamma$-ray band can be best interpreted by the leptonic models
\citep{Giom99,Tagl03}. The spectral energy distributions of blazars from radio to $\gamma$-ray energies have two characteristic components. In leptonic scenarios, the lower-energy part of photon SED can be produced by synchrotron emission, and the high energy part of SED
is produced by the inverse Compton (IC)
scattering 
\citep{Samb96,Tagl03}.
The source photons could be synchrotron photons (synchrotron self-Compton, SSC), or come from the outer radiation fields of jet (external Compton, EC) \citep[e.g.][]{Derm93,Derm09,Mars96,Blaz00}.
Our results
\textbf{indicate} that the 
variability timescales of \object{Mrk 421} at radio 15 GHz, X-ray and $\gamma$-ray are both consistent with each other, and the
variations of the three wavebands are correlated. This suggests that the radiations of three different \textbf{bands} are from the electrons with same property in the leptonic models. Namely, the emission of radio and X-ray is produced by the synchrotron radiation process, and the $\gamma$-ray is produced by IC scattering from the synchrotron photons. This is in agreement with the results obtained by many authors \citep[e.g.][]{Abdo14,Acci11,Alek12,Balo13,Gaur12,Hora09,Ushi10}, who suggest that emission of \object{Mrk 421} is dominated by the SSC processes.

In addition, our results also suggest that the emission of X-ray lag behind that of $\gamma$-ray with a timescale about 2 days, and the emission of X-ray and $\gamma$-ray lead that of radio with a timescale about 50 days. This time delay between different wavebands could be caused by the frequency stratification expected in the shock-in-jet models \citep{Gaur12,Mars96,Max14,Wagn95}. In this model, a moving radiation region, restricted to the jet, produces the radio to $\gamma$-ray activity, and the moving emission region moves outward at the bulk jet speed $\beta c$ \citep{Max14}. In the process of moving, the high frequency radiation would become observable earlier than the low frequency radiation, because the opacity region from the central engine of high frequency radiation is smaller than that of low frequency radiation. That is, the low frequencies variability lags behind the high frequencies variability is because the higher frequency emission originates from the upstream of the lower frequency emission \citep{Max14}.
Hence, the emission variability should appear in higher frequencies first, followed by the lower frequencies.
Thus, the emission of $\gamma$-ray would lead that of the X-ray band, and the X-ray variability would lead the radio band.

This work made use of
data supplied by the Owens Valley Radio Observatory 40 m monitoring programme, the UK Swift Science Data Centre at the University
of Leicester and Fermi Gamma-Ray Space Telescope project. This work
is supported by the National Natural Science Foundation of China
(11403015, U1531105, 11463001), and the Natural Science Foundation of Yunnan Province
(2012FD055, 2013FB063), and the Program for Innovative
Research Team (in Science and Technology) in University of Yunnan
Province (IRTSTYN), and the Young Teachers
Program of Yuxi Nurmal University. 



%


\clearpage

\begin{figure}
  \centering
   \includegraphics  [width=5.0in, angle=0]{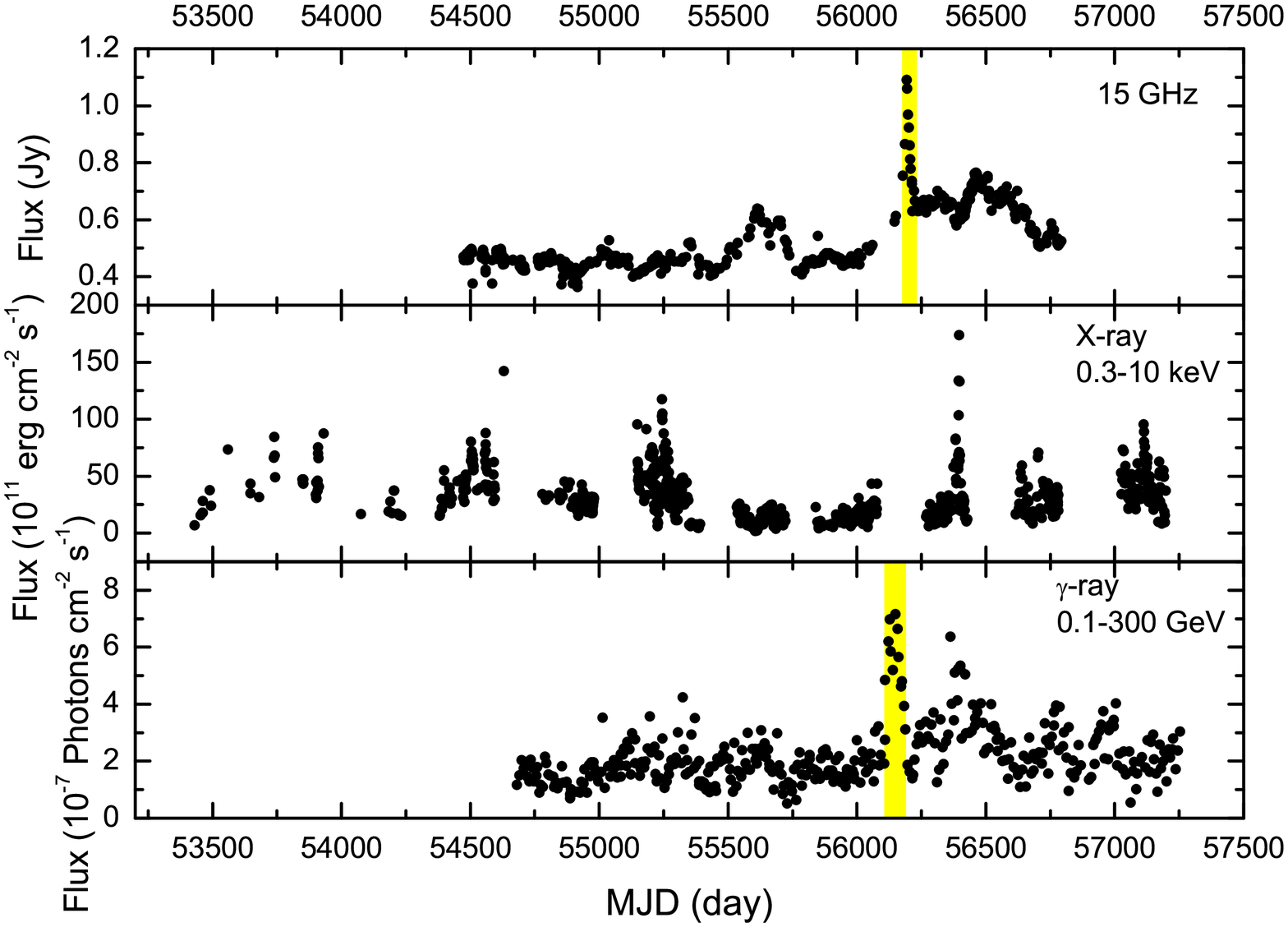}
  \caption{The light curves of Mrk 421 at 15 GHz, X-ray and $\gamma$-ray band. \textbf{In the 15 GHz and $\gamma$-ray panel, the shaded area is the 2012 flare. In X-ray band, there is a gap at the same time interval.}}
\end{figure}
\clearpage
\begin{figure}
  \centering
   \includegraphics  [width=5.0in, angle=0]{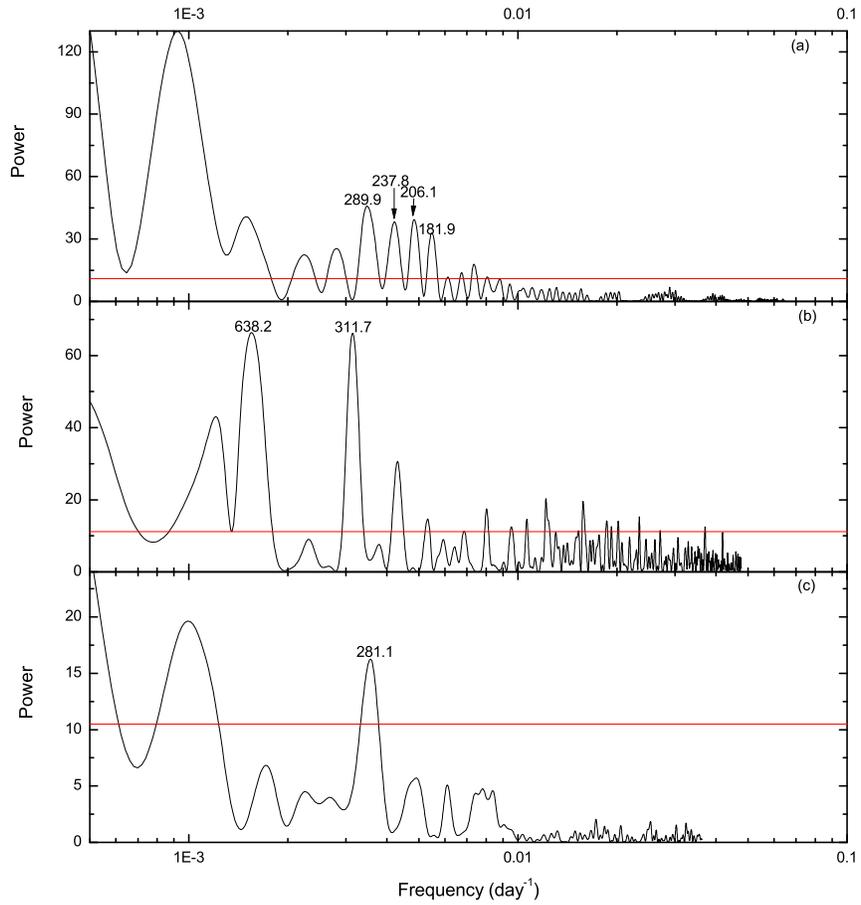}
  \caption{The Lomb-Scargle periodogram results of Mrk 421. The (a), (b) and (c) panel show the results of the 15 GHz, X-ray and $\gamma$-ray band light curves, respectively. The red lines indicate FAP levels equaling to 0.01.}
\end{figure}
\clearpage

\begin{figure}
  \centering
  \includegraphics  [width=6.0in, angle=0]{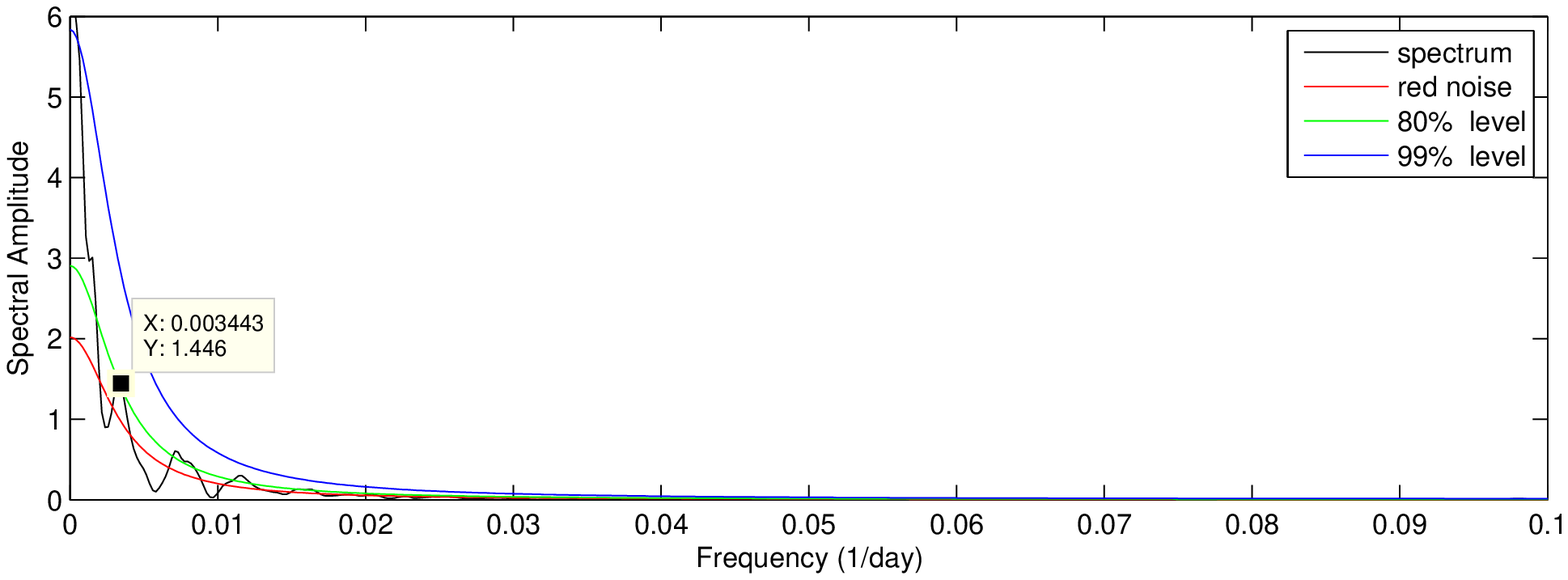}
  \hspace{0.005in}
  \includegraphics  [width=6.0in, angle=0]{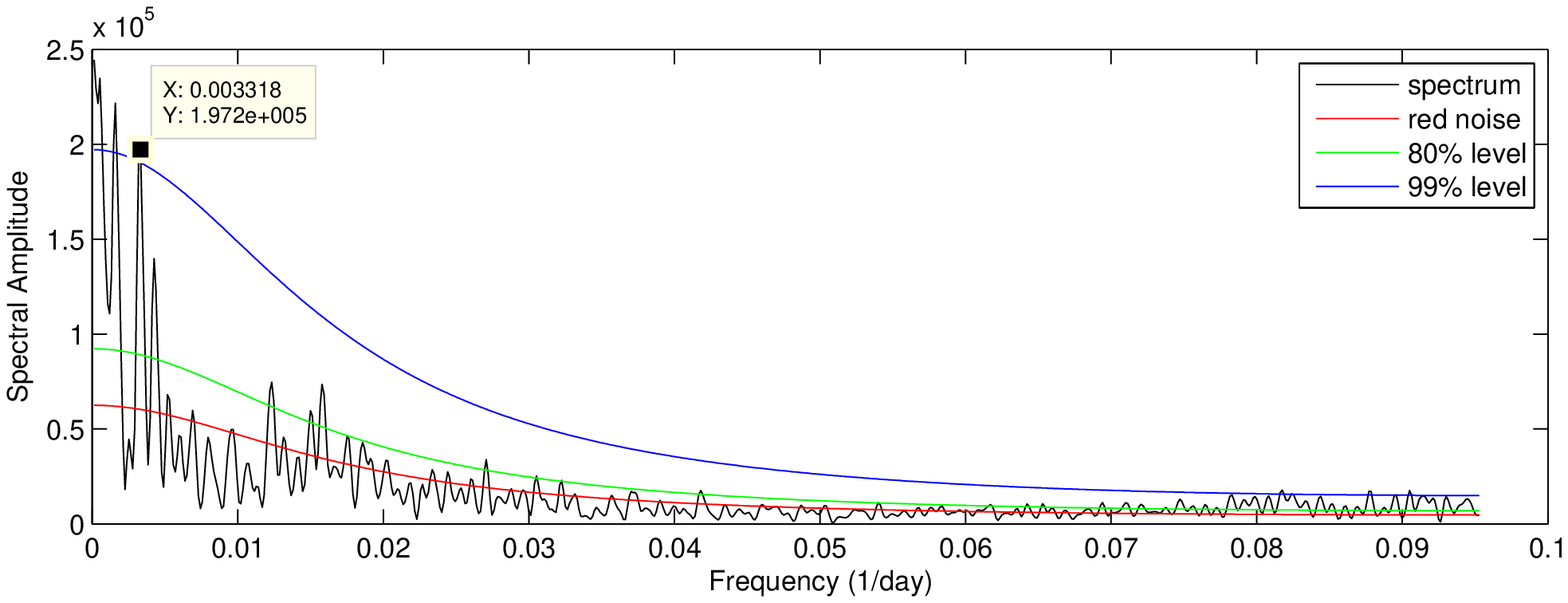}
  \hspace{0.005in}
  \includegraphics  [width=6.0in, angle=0]{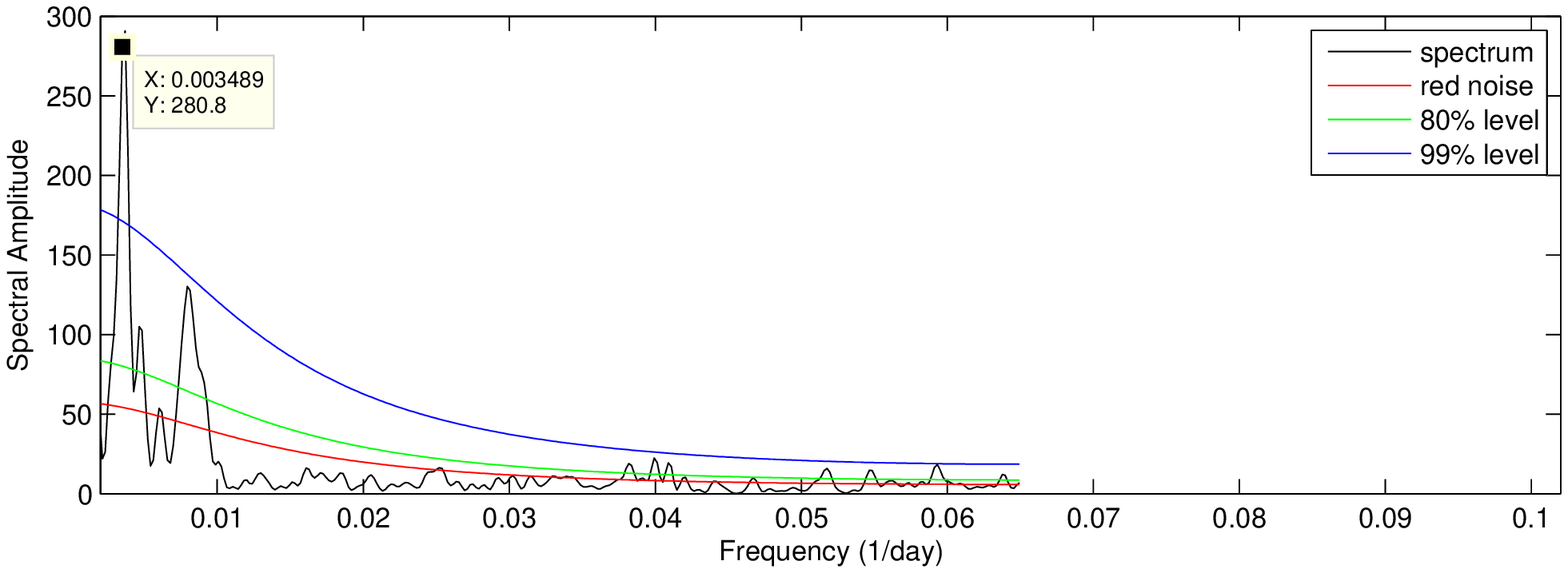}
  \hspace{0.005in}
\caption{The results of Mrk 421 calculated by the REDFIT38 software. The top, middle and bottom panel show the results of the 15 GHz, X-ray and $\gamma$-ray band light curves, respectively.}
\end{figure}
\clearpage
\begin{figure}
  \centering
   \includegraphics  [width=5.0in, angle=0]{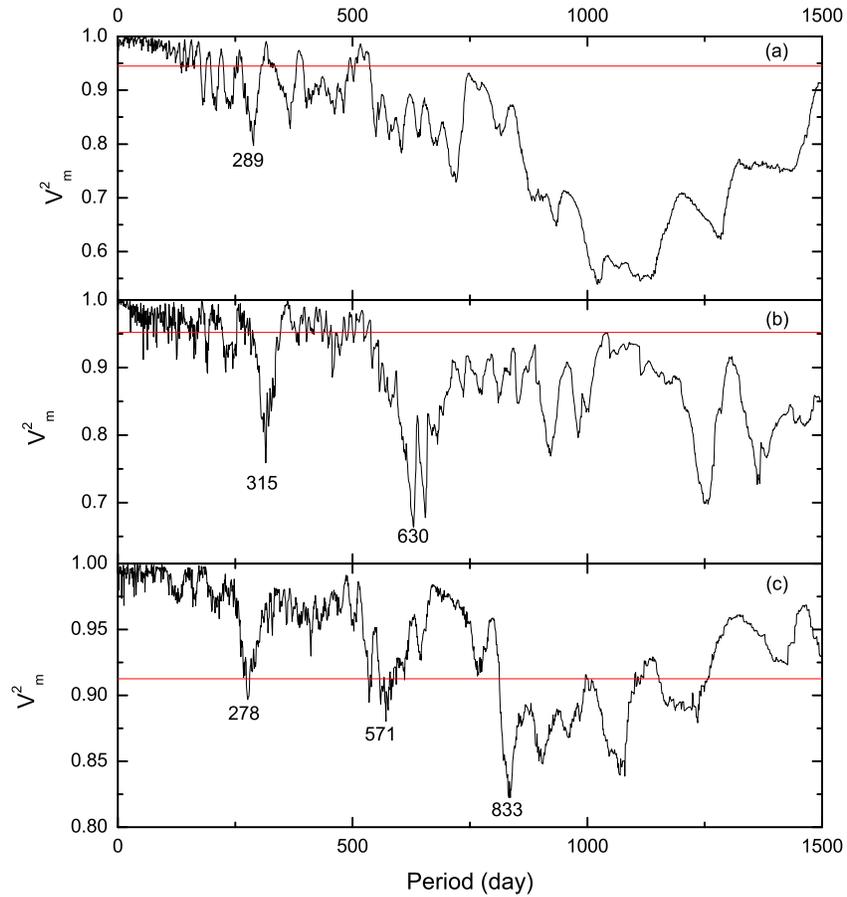}
  \caption{Period analysis results of Jurkevich method for Mrk 421. The (a), (b) and (c) panel show the results of the 15 GHz, X-ray and $\gamma$-ray band light curves, respectively. The red lines indicate FAP levels equaling to 0.01.}
\end{figure}

\clearpage
\begin{figure}
  \centering
  \includegraphics  [width=3.0in, angle=0]{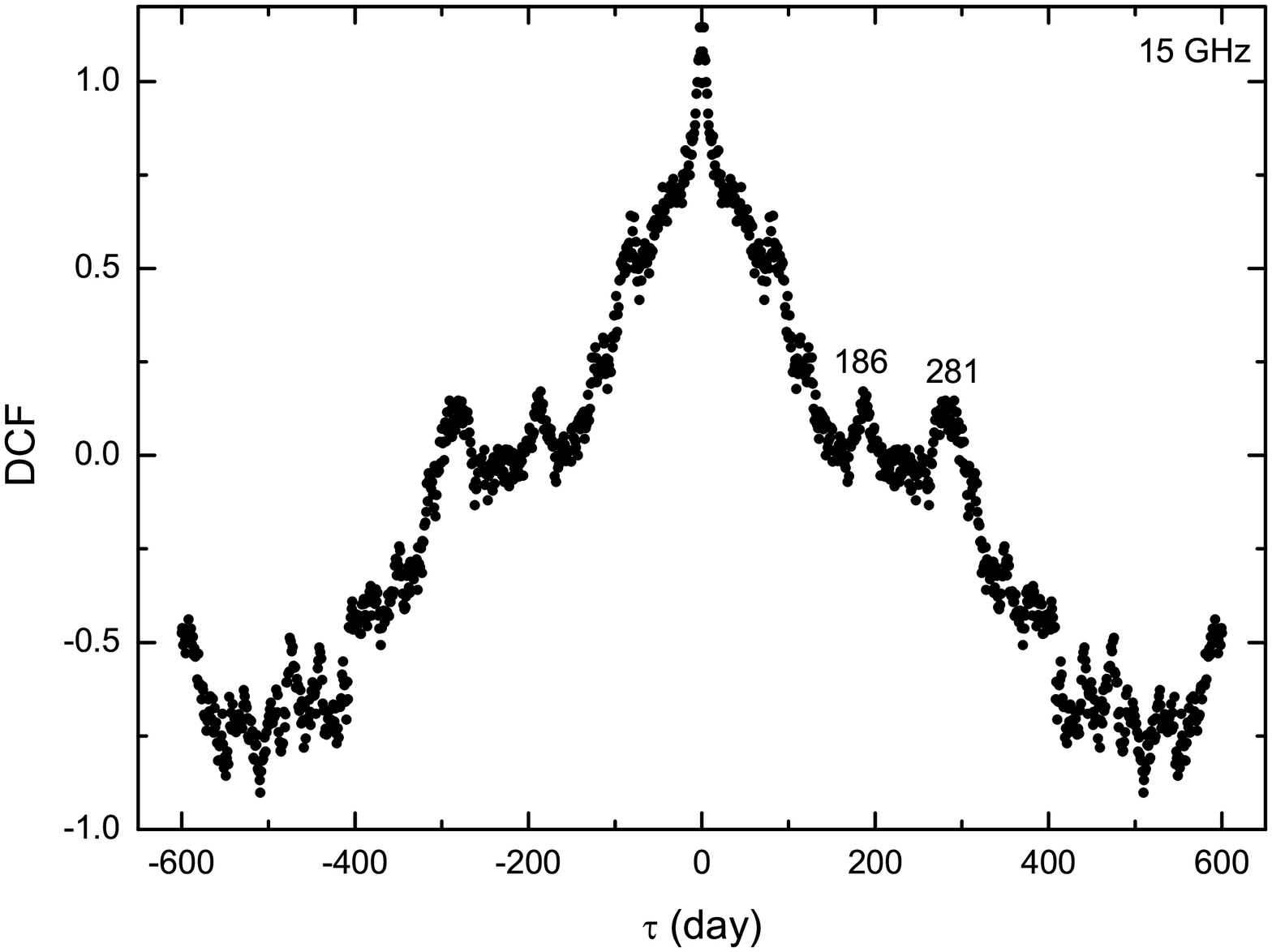}
  \hspace{0.005in}
  \includegraphics  [width=3.0in, angle=0]{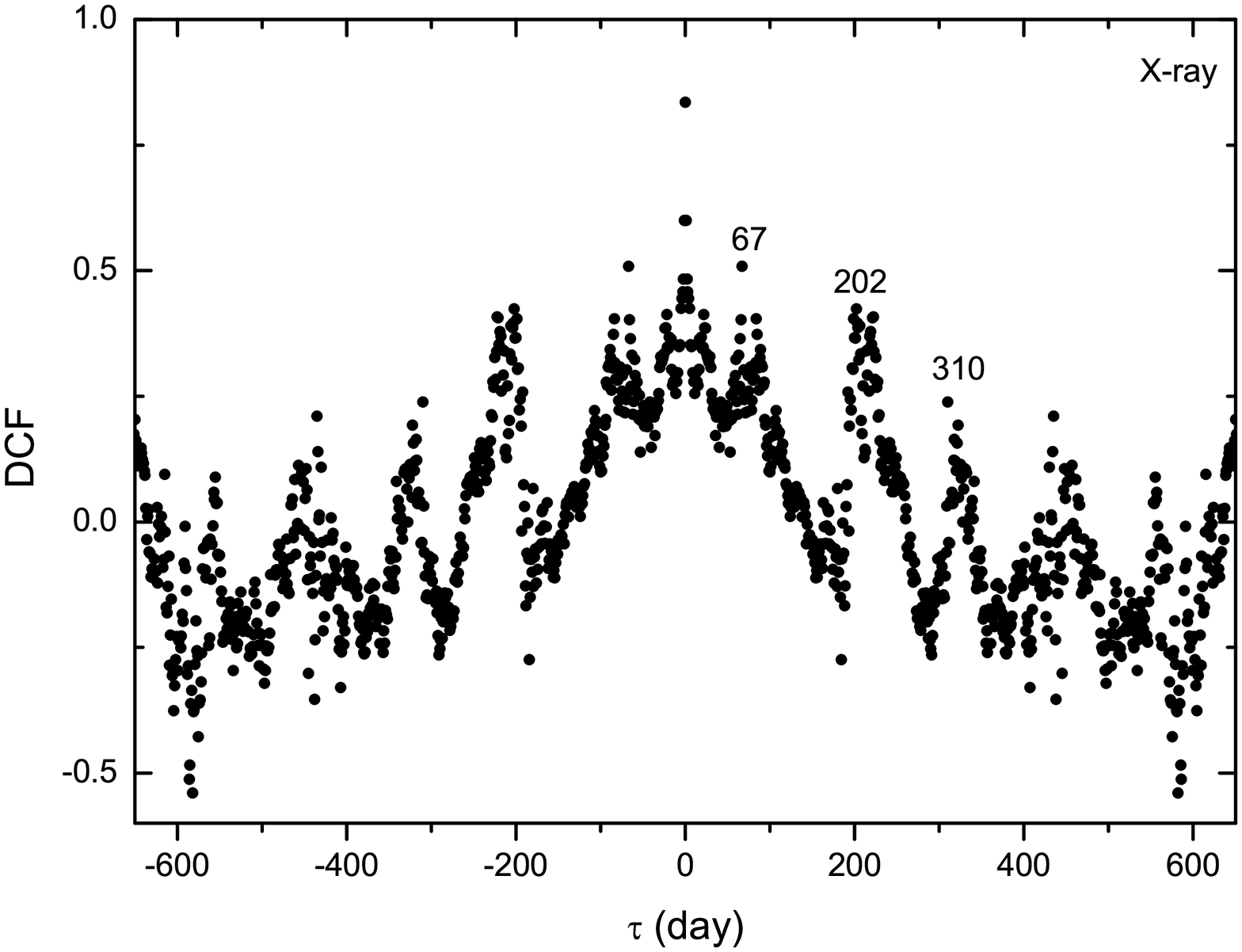}
  \hspace{0.005in}
  \includegraphics  [width=3.0in, angle=0]{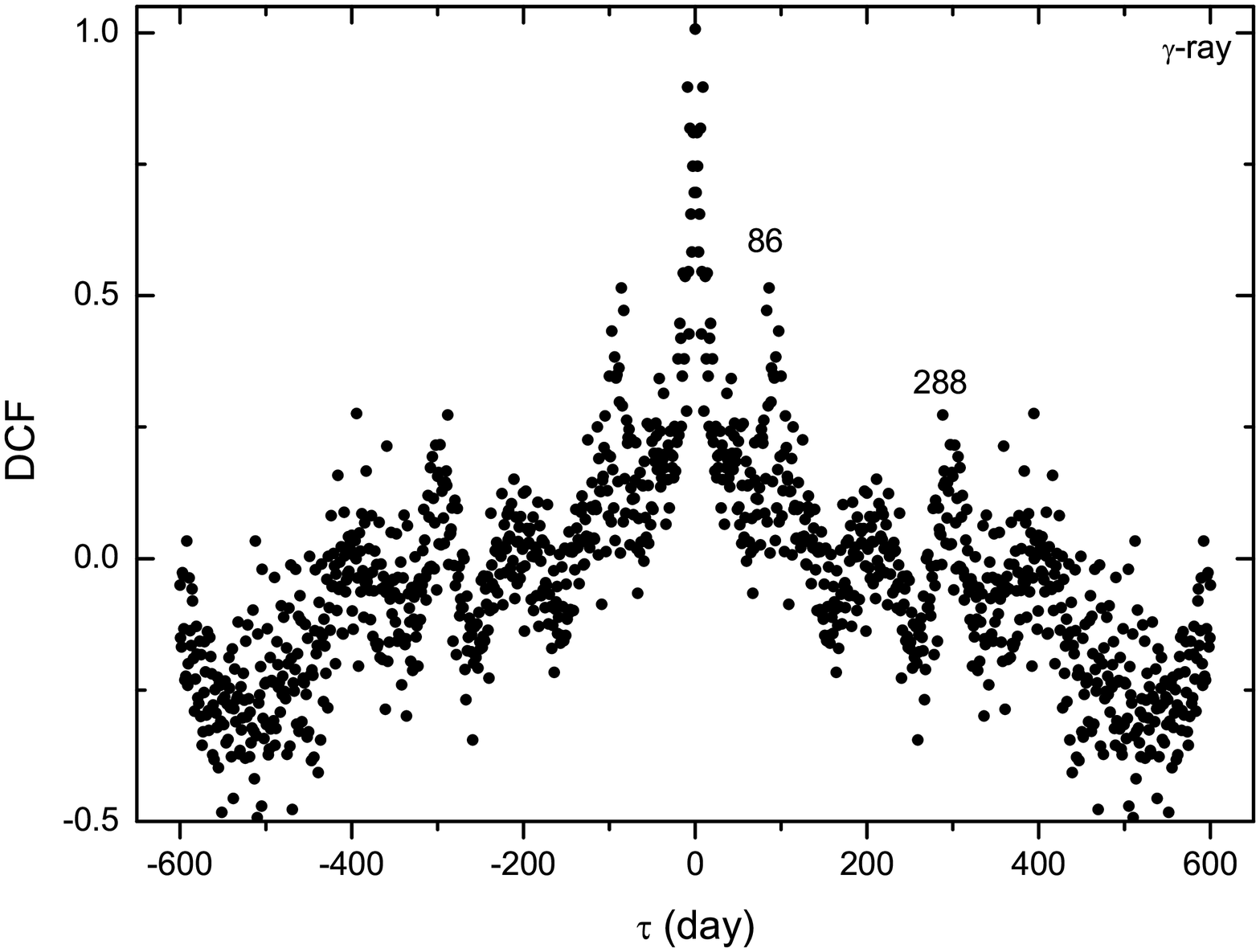}
  \hspace{0.005in}
\caption{The DCF results of Mrk 421. The (a), (b) and (c) panel show the results of the 15 GHz, X-ray and $\gamma$-ray band light curves, respectively.}
\end{figure}

\clearpage
\begin{figure}
  \centering
   \includegraphics  [width=6.5in, angle=0]{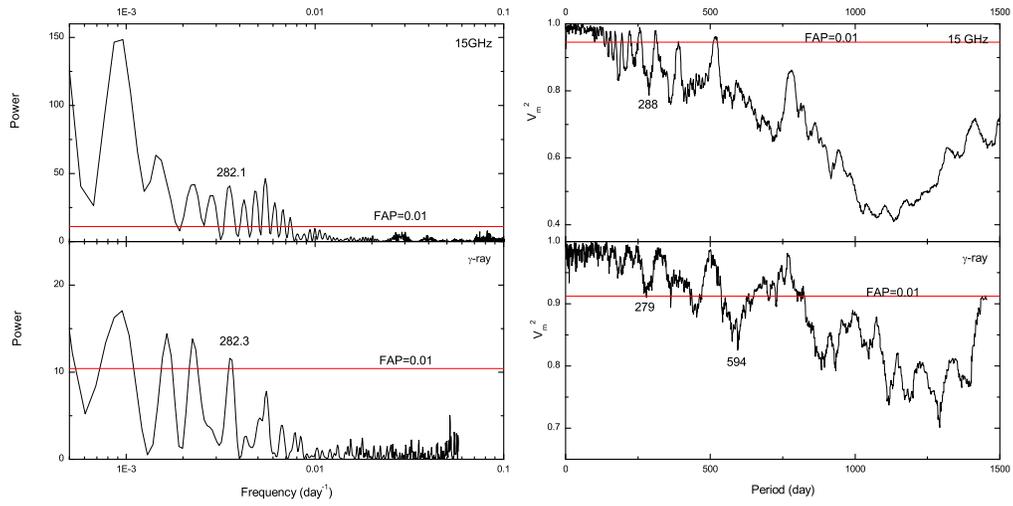}
  \caption{The left and right panel are the results calculated by the Lomb-Scargle periodogram and Jurkevich method using the data removing the 2012 flare data, respectively. The red lines indicate FAP levels equaling to 0.01.}
\end{figure}
\clearpage
\begin{figure}
  \centering
   \includegraphics  [width=5.0in, angle=0]{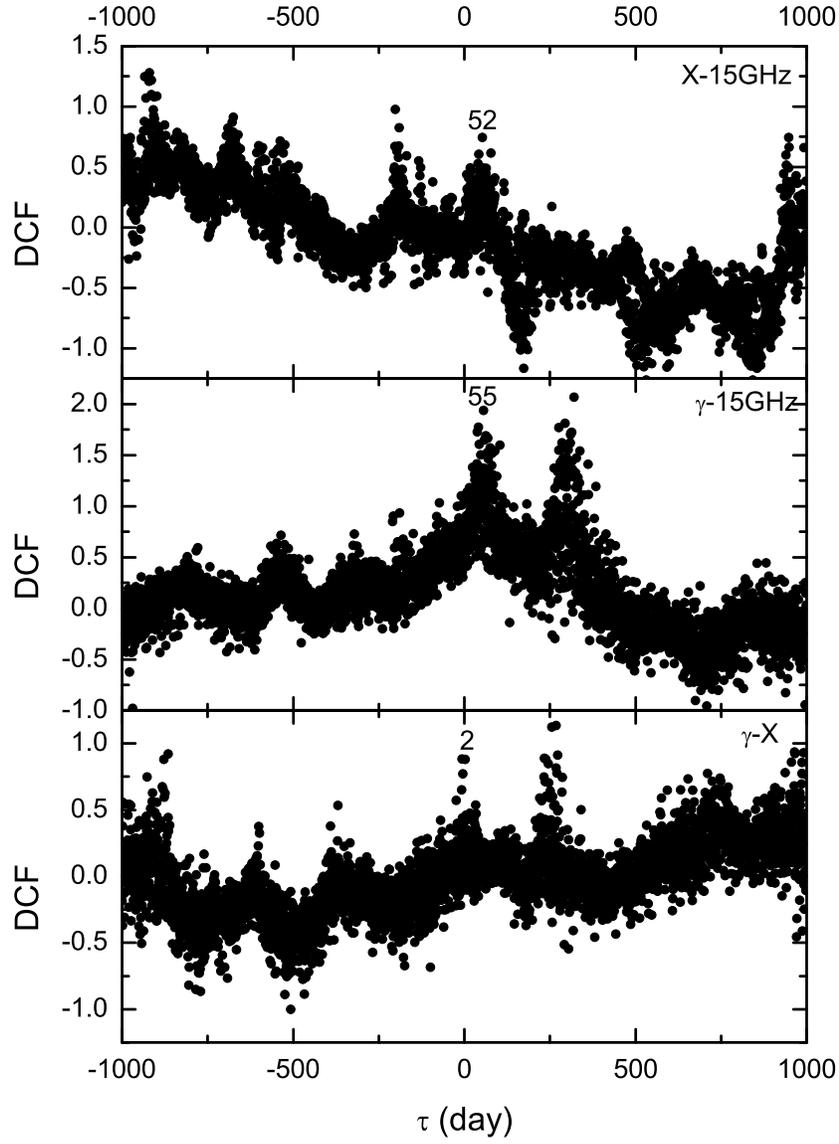}
  \caption{The correlations between the 15 GHz, X-ray, and $\gamma$-ray band of Mrk 421.}
\end{figure}

\begin{table}[!ht]
\centering
\caption[1]{Variability periods of Mrk 421 calculated by Lomb-Scargle periodogram, Jurkevich
method and DCF method.}
\centering
\begin{tabular}{c|cccc|ccc|ccc} \hline
  & \multicolumn{4}{c|} {15 GHz} & \multicolumn{3}{c|}{X-ray} &  \multicolumn{3}{c}{$\gamma$-ray} \\ \hline
  Power (days)& 181.9&206.1&237.8&289.9   & 311.7& \multicolumn{2}{c|}{638.2}  &  \multicolumn{3}{c}{281.1} \\ \hline
  REDFIT38 (days) &  \multicolumn{4}{c|}  {290.4}  &    \multicolumn{3}{c|}{301.4}    &  \multicolumn{3}{c}{286.6}       \\ \hline
  Jurkevich (days) &  \multicolumn{4}{c|}  {289}  &    315& \multicolumn{2}{c|}{630}    &  278&571&833       \\ \hline
  DCF (days)&  \multicolumn{2}{c} {186}&\multicolumn{2}{c|}{281}&   67&202&310 &  86& \multicolumn{2}{c}{288} \\ \hline
  Period(days)& \multicolumn{4}{c|} {281-290.4} & \multicolumn{3}{c|}{301.4-315} &  \multicolumn{3}{c}{278-288} \\ \hline

\end{tabular}
\end{table}
\clearpage

\end{document}